\documentclass[a4paper]{jpconf}
\usepackage{graphicx}
\usepackage{natbib}
\bibpunct{(}{)}{;}{a}{}{,} 
\usepackage{txfonts}
\newcommand{\degree}{\ensuremath{^\circ}}
\newcommand{\Vhelio}{V$_{helio}$ }

\begin{document}

\title{A new perspective on the interstellar cloud surrounding the Sun from UV absorption line results}

\author{Cecile Gry$^1$ and Edward B. Jenkins$^2$}

\address{$^1$ Aix Marseille Universit\'e, CNRS, LAM (Laboratoire d'Astrophysique de Marseille) UMR 7326, 13388, Marseille, France}
\address{$^2$ Department of Astrophysical Sciences, Princeton University Observatory, Princeton, NJ 08544, USA }

\ead{cecile.gry@lam.fr, ebj@astro.princeton.edu}

\begin{abstract}
We offer a new, more inclusive, picture of the local interstellar medium, where it is composed of a single, 
monolithic cloud that surrounds the Sun in all directions. Our study of velocities based on  Mg~II and  Fe~II ultraviolet absorption lines indicates that the cloud has an average motion  consistent with the velocity vector of gas impacting the heliosphere and does not behave like a rigid body: gas within the cloud is being differentially decelerated in the direction of motion, and the cloud is expanding in directions perpendicular to this flow, much like the squashing of a balloon. 
The outer boundary of the cloud is in average 10 pc away from us but is highly irregular, being only a few parsecs away in some directions,  with possibly a few  extensions up to 20 pc. 
Average H I volume densities vary between 0.03 and 0.1 cm-3 over different sight lines.
Metals appear to be significantly depleted onto grains, and there is a steady increase in this effect from the rear of the cloud to the apex of motion.  There is no evidence that changes in the ionizing radiation influence the apparent abundances. 
Additional, secondary velocity components are detected in 60\% of the sight lines. Almost all of them appear to be interior to the volume holding the gas that we identify with the main cloud. 
Half of the sight lines exhibit a secondary component moving at 
about - 7.2 km/s
with respect to the main component, which may be the signature of an implosive shock propagating toward the cloud's interior.  

\end{abstract}

\section{Introduction}
Our aim is to propose a simplified picture of the local interstellar medium (LISM), where it is composed mainly of a single, continuous cloud, which envelops the Sun in all directions.  This outlook differs from the current, commonly accepted view that the LISM is a collection of identifiable cloud structures, all spatially and kinematically distinct from one another, participating in a common bulk motion (cf Frisch, Redfield \& Slavin, 2011 for a review). In the most complete version of this model, Redfield \& Linsky (2008) interpreted the different velocity components detected in absorption in the spectra of nearby stars in terms of 15 coherently moving clouds, all moving at different but similar velocities. In their picture  the heliosphere is not immersed in the local interstellar cloud (LIC), but  is situated near its edge, in the transition region between  the LIC and another cloud referred as the  G cloud. With such a configuration, whereby the Sun is surrounded by several distinct and separate clouds, we would expect to find some lines of sight going through the gaps between the clouds, which should create at least a few cases showing no absorption.  However such a line of sight has never been found: every stellar spectrum including a sensitive enough absorption line in the UV shows at least one interstellar  component. We note also that three of the clouds (LIC, G and a third one denoted NGP for North Galactic Pole) almost complement each other  to cover most of the sky with no overlap, and are described with  similar velocity vectors. 
A question arises: Do the data support a picture that these clouds could be unified into a single cloud?  For instance, the G cloud had been introduced by Lallement \& Bertin (1992) to account for a slight velocity discrepancy in the direction of $\alpha$ Cen. Now that we have a larger sample of nearby sight lines and a better knowledge of the ISM flow in the heliosphere, we can revisit the question,  ``Is the G cloud still needed as a distinct cloud?"
We therefore decided to re-examine the LISM UV-absorption database with a new focus, i.e. making the assumption that the ISM close to the solar system can be viewed as a unique, continuous medium that envelops  the Sun in all directions and, in the simplest approximation, produces one
absorption component in every line of sight. With this assumption in mind, 
we examine 
how this conforms with a picture of a homogeneous cloud, moving with a coherent velocity vector as a rigid body.     
\section{The data}
We re-examine the Mg II and Fe II results from UV lines observed with HST (GHRS and STIS) at high resolution (R$\simeq100\, 000$, or $\Delta V\simeq3\,{\rm km~s}^{-1}$). 
We use the data compiled in Redfield \& Linsky (2002) Table 3 (for Mg II) and Table 4 (for Fe II), 
providing the velocity, broadening parameter, and column density for each component.
To guarantee the homogeneity in sensitivity and preserve the integrity of the velocity calibration, we do not include data from ground-based spectroscopy; therefore the CaII or NaI results are not included in this analysis. Note that we do not reanalyze the data; we adhere to the published results listed by Redfield \& Linsky (2002). However we have checked the velocity scales for data that had new wavelength calibrations, resulting in 6 slight velocity changes.
\\
Our resulting database includes 111 components, in a total of 59 lines of sight, of which 54 are observed in Mg II, 32 are observed in Fe II, and 27 are observed in both elements.
New observations in additional directions were reported at this conference (Redfield et al, this volume) and  appeared later in a paper by Malamut et al. (2014). The future addition of these new results should help to refine the conclusions presented here.
\section{Identification of a single circum-heliospheric cloud}
\subsection{Step 1: identification of the component most likely to originate in the circum-heliospheric cloud}
In every line of sight (l$_i$,b$_i$) we select the absorption component closest in velocity to  V$_{i,helio}$, i.e. the projection onto the sight line of the velocity vector of the ISM flow interacting with the heliosphere as measured by Ulysses and previous experiments (M\"obius et al. 2004) or by IBEX (McComas et al. 2012). We call this component "Component~1". There is by definition one Component~1 in every line of sight.
\subsection{Step 2: Mean velocity vector that best fit all Components 1}
We perform a least squares fitting of the Components~1 sample to a single vector in the sky in the heliocentric reference frame. The best-fit gives the following mean vector: 
\begin{figure}[h]
\includegraphics[width=16pc]{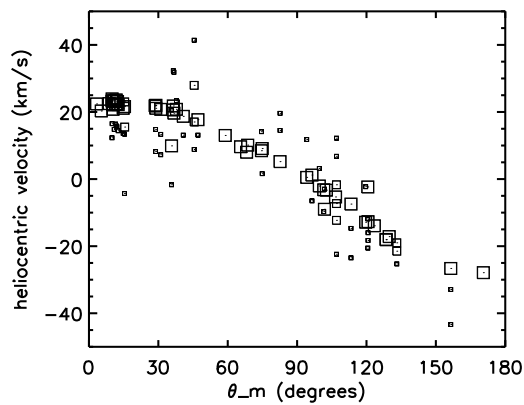}\hspace{2pc}%
\begin{minipage}[b]{20pc}\caption{Measured velocities of all components plotted against  $\theta_{m}$, the angular distance to the apex of the Components~1 mean vector. Large symbols indicate components that are dominant in their line of sight in terms of column density. Medium-size symbols depict cases where two components in a line of sight have identical column densities within their error bars. The smallest symbols indicate components that are clearly secondary.  The adopted reference column densities are those of Fe~II, when available, and Mg~II otherwise. \label{fig:velocity-domine} }
\end{minipage}
\end{figure}
\[l_m=185.84\pm0.83\degree ; \,b_m=-12.79\pm0.67\degree ;  V_m=25.53\pm0.26 {\rm km/s}\]
Our measured velocity vector is indeed very close to both solar system measurements: it has the same direction as the IBEX vector, but the velocity amplitude is closer to that of the Ulysses measurement.
We define for each line of sight the angle $\theta_m$,  representing its angular distance to the apex of the above mean velocity vector. 
On Figure~\ref{fig:velocity-domine} our Components~1 stand out as forming a conspicuous V$_m\cos\theta_{i,m}$ curve, where V$_m=25.53\,{\rm km~s}^{-1}$. They also appear to be the dominant component (shown as large squares) in most sight lines. 
\subsection{Step 3:  Components~1 velocity deviations from the best-fit velocity vector}
\begin{figure}[h]
\begin{minipage}{18pc}
\includegraphics[width=14pc]{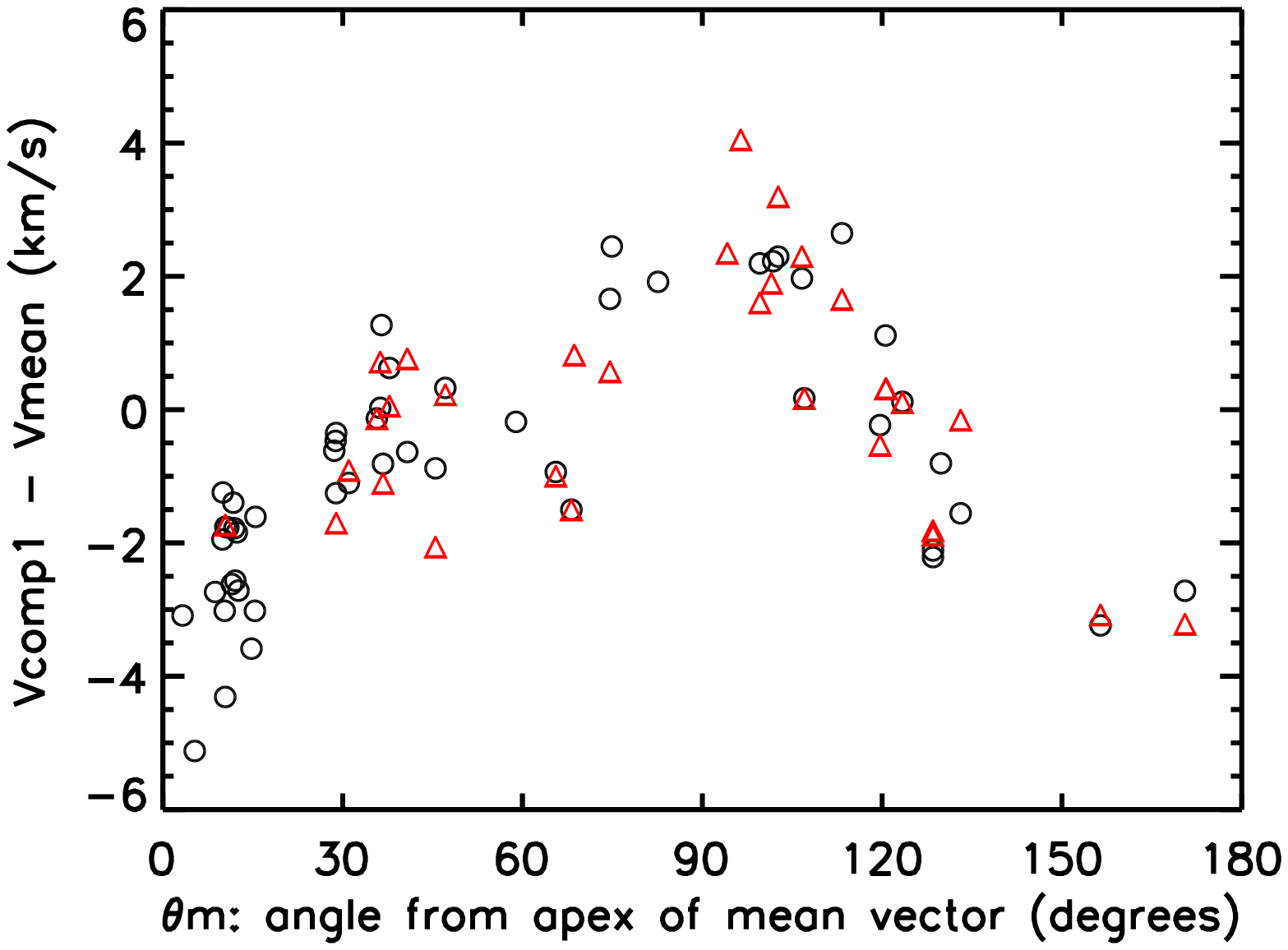}
\caption{Velocity residuals of Components~1, the components closest to \Vhelio, relative to the V$_m$cos$\theta_{i,m}$ relationship for the best-fit vector derived from the heliocentric Mg~II velocities. Circles: Mg~II, triangles: Fe~II
\label{fig:VMgII} }
\end{minipage}\hspace{2pc}%
\begin{minipage}{18pc}
\includegraphics[width=14pc]{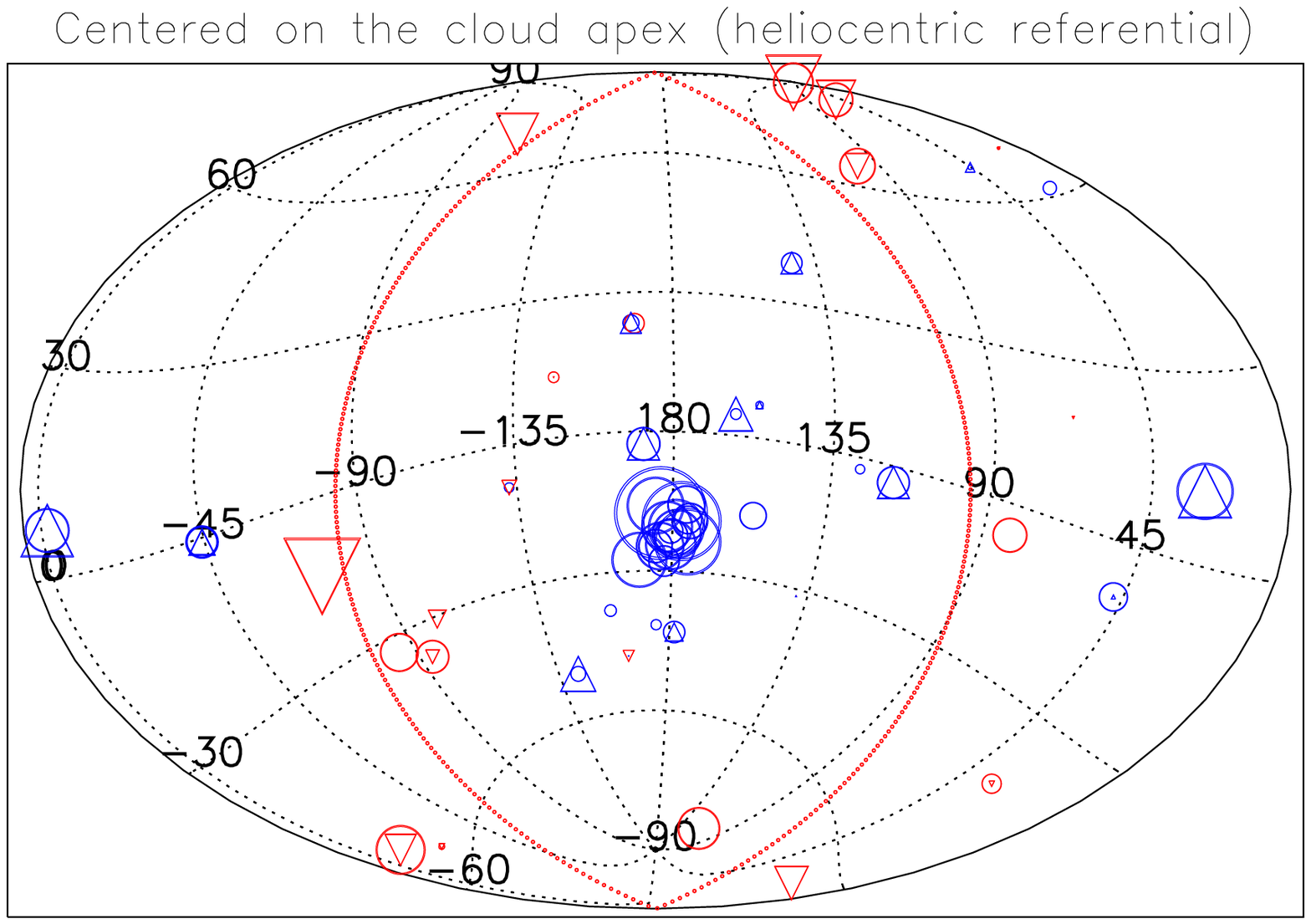}
\caption{Components~1 velocity deviations from the best-fit vector, in Galactic coordinates but centered on the apex of motion.  Red/blue: positive/negative deviations. Symbol size: magnitude of the deviations. Red curve: circle 90\degree\ away from the direction of motion. \label{fig:resid-aitoff} }
\end{minipage} 
\end{figure}
\noindent Figure~\ref{fig:VMgII} exhibits a clear trend of the velocity deviations with $\theta_m$: they are negative  toward the apex and anti-apex directions and positive in directions perpendicular to the motion.
When the trend is taken into account as a systematic velocity gradient, the remaining dispersion is comparable to the velocity determination uncertainties ($\simeq$ 1.1 km s$^{-1}$). Figure~\ref{fig:resid-aitoff} shows that the large positive deviations (red symbols) are indeed roughly distributed along a great circle at 90\degree away from the direction of motion, showing that the expansion perpendicular to the flow vector is seen in all azimuthal angles. 
\subsection{Conclusion: a single cloud}
We think that the simple, symmetrical, single hump pattern exhibited in Fig.~\ref{fig:VMgII}, and the coherent spatial distribution of the residuals, 
indicate this is a fundamental, second-order dynamical effect associated with an otherwise coherently moving cloud. We conclude that, provided we relax the assumption that the cloud is rigid, our Components~1 sample can indeed be interpreted as the projected velocities of a unique interstellar cloud, detected in all directions, with a mean velocity close to the ISM flow into the heliosphere: the Local Cloud. This cloud accounts for 53 \% of the components in number. It accounts for 70 \% of the column density in the first 50 pc. It includes most of the LIC, G, NGP, Blue, Leo, Aur, Cet, clouds and half of the Eri and Gem clouds components in the Redfield \& Linsky (2008) model.
\section{Characteristics of the cloud}
\subsection{Dynamics within the cloud}
We developed a simple, idealized model for the velocity deviations. We start with a spherical volume centered on the observer and estimate the velocity perturbations when the sphere is differentially decelerated. To maintain its original volume (and internal pressure) the cloud is deformed into an oblate ellipsoid that has its minor axis aligned with the deceleration vector. We derive the velocity perturbations $\Delta$V($\theta_d$) away from a solid-body motion, expressed as a function of $\theta_d$, which is the angle away from the minor axis of the ellipsoid:
\centerline{$\Delta {\rm v}(\theta_d)=C\,\left[ (1-e^2)^{-1/6}(1-e^2\cos^2\theta_d)^{1/2}-1\right]~,$}
where $e$ is the eccentricity of the ellipsoid and  $C$ is a constant. Provided the variable $e$ is not large ($e<0.8$), the shape of the curve tracing $\Delta {\rm v}$ as a function of $\theta_d$ is invariant and its amplitude is governed only by the product of $C$ and $e$. \\
Minimizing the quadratic differences of the observed 
velocity deviations with respect to $\Delta$V($\theta_d$),
we derive the direction of the minor axis (i.e. the origin of the angle $\theta_d$):
\centerline{$l_d=174 \pm5 \degree; \,\,\,b_d=-12\pm4\degree~,$}
which is relatively close to the apex of the mean velocity vector found in the previous section. If the differential deceleration and distortions are due to the cloud interaction with an external medium or a magnetic field,  the minor axis found for the spatial deformation  is very likely to indicate a fundamental direction for the cloud's internal motions in response to these external forces, independent of the chosen reference frame. For this reason, we adopt ($l_d, b_d$) as the reference direction for the cloud.
\subsection{Extent and density of the cloud}
\begin{figure}[!htbp]
\includegraphics[width=35pc]{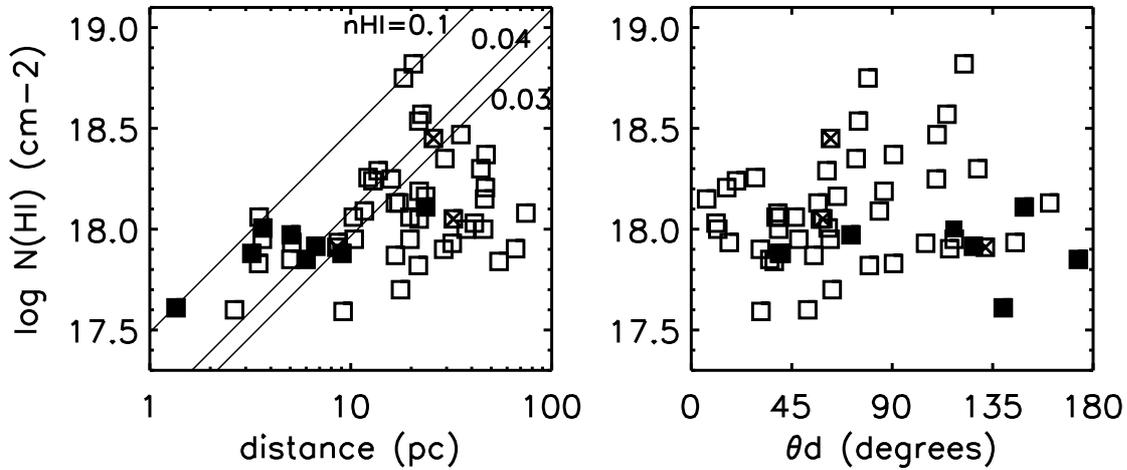}
\caption{
 H~I column densities. \opensquare: Components~1. \fullsquare: targets presenting evidence for an astrosphere and only one H~I component, i.e. sight lines completely filled with the Local Cloud. Crossed  \opensquare: targets where the presence of the astrosphere  is uncertain. Diagonal lines show lines of constant densities $n({\rm H~I})=0.03,\,0.04,\,0.1\,{\rm cm}^{-3}$ . 
\label{fig:distanghi} }
\end{figure}

Mg II and Fe II column densities exhibit no correlation  with distance: this means that the cloud is small compared to target distances. On the other hand, the column densities show a high dispersion and there is no obvious trend with direction: we conclude that the cloud is very irregular in shape.\\
To study the extent and density of the cloud we examine the H I compilation and results from Wood et al (2005), with particular attention to their detections of astrospheres. Since an astrosphere arises from the interaction of the stellar wind with neutral atoms, its presence for a particular line of sight means that the target star is embedded in neutral gas. 
In the H I sample an astrosphere is detected in 8  single-component lines of sight, where we conclude that  the Local Cloud fills up the entire sight line, and where the ratio N(HI)/d reliably measures the average density. The latter  varies in these sight lines from 0.03 to 0.1 cm$^{-3}$, yielding a mean neutral density in the Local Cloud of 0.049 cm$^{-3}$, in agreement with the solar system measurements of interstellar pick-up ions and anomalous cosmic rays of n(HI)= 0.055 $\pm$ 0.029 cm$^{-3}$ (Gloeckler et al. 2009). Inspection of Figure ~\ref{fig:distanghi} shows that (1) the Local Cloud extends up to 20 pc in a few sight-lines and to less than 15pc in most of them, 
(2) if we adopt n(HI)~=~0.05 cm$^{-3}$, the mean extent of the cloud is $<$N(HI)/n(HI)$>$ = 9.3 $\pm$ 7.6 pc, (3) in no directions does the Sun lie near the cloud edge: the minimum distance to the boundary is somewhere between 1.3 and 4.7 pc, depending on the density value chosen, and (4) highest column densities are found in directions perpendicular to the cloud deformation minor axis.
\subsection{MgII and FeII abundances}
\begin{figure}[!htbp]
\includegraphics[width=24pc]{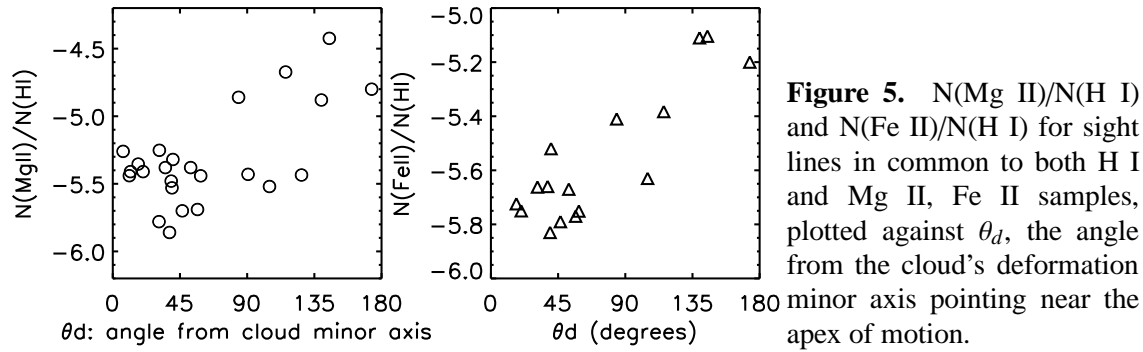}
\begin{minipage}[b]{11pc}
\caption{N(Mg~II)/N(H~I) and N(Fe~II)/N(H~I) for sight lines in common to both H~I and Mg~II, Fe~II samples, plotted against $\theta_d$, the angle from the cloud's deformation minor axis pointing near the apex of motion.
\label{fig:abundance} }
\end{minipage}
\end{figure}

As illustrated in Figure~\ref{fig:abundance} a correlation exists between the abundances of MgII and FeII and the angle $\theta_d$, which suggests the presence of an abundance gradient within the cloud.
\\
We first check if the observed gradient can be explained by ionization effects. The LISM is known to be partially ionized and  its ionization is maintained by a strong UV flux dominated by the two stars $\epsilon$ and $\beta$ CMa, located respectively at $\theta_d$ = 65\degree and 51\degree. Shielding by the cloud material makes the ionizing flux  decrease with cloud depth and this means that the ionization should be maximum in the direction of the two stars and decrease away from it. However, there is no visible correlation of the Mg~II and Fe~II abundances with the angular distance to the star $\epsilon$ CMa. On the other hand, in regions were photons more energetic than 13.6 eV can penetrate the cloud, an element can be more or less ionized than H, depending on its relative photoionization to recombination rate ratio $\Gamma$(X)/$\alpha$(X) compared to $\Gamma$(H)/$\alpha$(H) (Jenkins 2004). As a consequence, we would expect to find Fe~II more ionized and Mg~II less ionized than H, making N(Fe~II)/N(H~I) lower and N(Mg~II)/N(H~I) higher than in fully neutral gas. However, we will see that the opposite is true since Mg~II appears more depleted than expected. Therefore there is no evidence that ionization influences the apparent abundances.

The trends displayed in Figure~\ref{fig:abundance} are thus most probably related to the gas-phase abundances of the two elements. The magnitude of the depletion is gradually increasing from the upwind ($\theta_d=180\degree$) to the downwind direction ($\theta_d=0\degree$). This had been noticed already by Redfield \& Linsky (2008) and interpreted as a physical difference between the LIC and the G cloud, whereas in our picture it is the sign of an abundance gradient within the cloud. The metal abundances are roughly 4 times (0.6 dex) higher in the rear of the cloud than in the front. We note that there is a constant depletion ratio of $\simeq$ 1.5 (0.17 dex) between Mg and Fe, independent of $\theta_d$, which is significantly lower that the mean interstellar value. 
The fact that we observe more depletion near the head of the cloud, compared to that at the rear (i.e., upwind direction), indicates  that  a progressive erosion of the grain population may either be taking place, perhaps as a consequence of whatever process is causing the differential deceleration, or may alternatively have occurred  in relation to the origin of the cloud as proposed by Kimura et al (2003). 
\section{Nature of the other components}
\begin{figure}[!htbp]
\includegraphics[width=18pc]{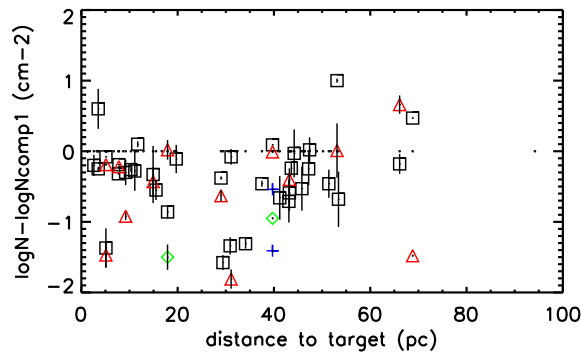}
\begin{minipage}[b]{19.5pc}
\caption{Column density ratios (in log) between the other components and their corresponding Component~1: log N/Ncomp1 against distance. 
The horizontally arranged dots are components 1. They show the distances of all lines of sight. The other components, numbered in order of increasing $|\Delta V|$ relative to Component~1 are as follows: Black squares are components 2, red triangles components 3, green diamonds components 4 and blue crosses components 5 and 6. 
\label{fig:othercomp-N} }
\end{minipage}
\end{figure}

\begin{figure}[!htbp]
\includegraphics[width=24pc]{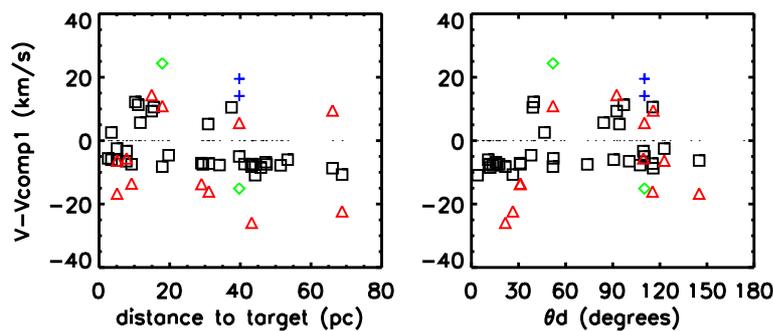} 
\begin{minipage}[b]{13.5pc}
\caption{Velocity shifts of the other components relative to Components~1,  against  distance and $\theta_d$.  
The meanings of different symbols are the same as those in figure~\ref{fig:othercomp-N}. Notice the preponderance of horizontally aligned symbols just below zero: these are what we refer to as the "Cetus Ripple" components.
\label{fig:othercomp-vdif} }
\end{minipage}
\end{figure}

\begin{figure}[!htbp]
\includegraphics[width=17pc]{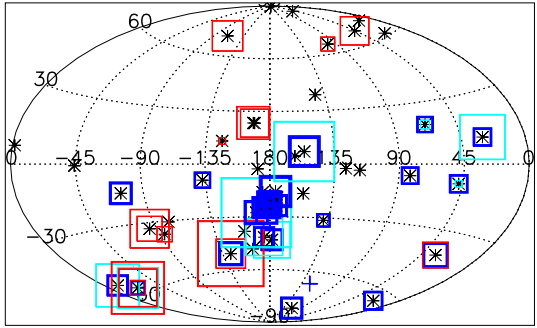} 
\begin{minipage}[b]{19.5pc}
\caption{Velocity shifts of the other components relative to Component~1. Black asterisks: all targets.  Squares: components other than Components~1. Red/blue:  positive/negative shifts.   Dark blue:  Cetus Ripple components.
Symbol size: amplitude of the shift. The blue cross indicates the rough center of the hemisphere occupied by the Cetus Ripple  components. 
\label{fig:aitoff}}
\end{minipage}
\end{figure}

Out of 59 sight lines, 34 (58\%) show at least one other component in addition to Component~1, 13 (22\%) show at least two other components, one shows three and another shows five additional components. We examine the characteristics of these other components to try and pinpoint their nature.
%
\\
\\
{\it 5.1. Secondary nature of the other components.}\\
Figure~\ref{fig:othercomp-N} shows that in any line of sight Component~1, i.e., the component whose velocity is most consistent with the velocity vector of gas impacting the heliosphere, is generally also the dominant component. 
Out of the 4 lines of sight  where another component is larger than Component~1, three of them  ($\iota$ Cap, $\upsilon$ Peg and G191B2B) are longer than 50 pc, meaning that the larger component may be situated far away. For the one short sightline toward 61 Cyg A (3.5 pc), the profile fitting is very uncertain and could in fact result in a reversed column density ratio for the two components if different b-values were chosen (Wood \& Linsky 1998). Therefore the components we have kinematically identified as the Local Cloud represent the bulk of the matter in the first 50 pc.
%
\\ 
{\it 5.2. Proximity of the other components.}\\
We note in Figure ~\ref{fig:othercomp-vdif} that 1) longer sight lines do not show more components than very short ones and 2) even the shortest ones exhibit multiple components: 5 sight lines shorter than 5 pc show 2 or  even 3 components. This implies that 
most of the secondary components exist at distances  short of those to the edges of the cloud, and that some of them could be situated within the same volume as that occupied by the local cloud.
%
\\
{\it 5.3. Consistency in the velocity shifts of the other components relative to the Local Cloud}\\
Figure ~\ref{fig:othercomp-vdif} also illustrates the fact that half of the secondary components have a consistent $\Delta V = -7.2\,\pm\,1.5{\rm (rms)~km~ s}^{-1}$ relative to the Local Cloud component and about half of the sight lines include such a component. Figure~\ref{fig:aitoff} shows that the components exhibiting the $-7.2\,{\rm km~s}^{-1}$ shift also cover approximately half of the sky. 
\noindent When fitting these components as a cloud having a coherent motion relative to the Local Cloud, we get a velocity vector originating from Cetus (l$_{CR}=124\degree\pm$4\degree, b$_{CR}=-67\degree\pm$3\degree), which is what motivated us to name them the Cetus Ripple components. However, their velocity shifts plotted against the angular distance to the best-vector origin do not follow the expected cosine relationship, but appear instead very constant over $\theta$. This shows that the Cetus Ripple components do not present a coherent linear motion, but instead we conclude that their motion is  characteristic of an implosion converging near our location.
%
\\
{\it 5.4. A plausible model for the Cetus Ripple as a shock progressing toward the cloud's interior}\\
Considering all known  characteristics of the pre-shock gas (i.e. the Local Cloud), we show that it is possible to define a shock velocity (between 20 and 26 ${\rm km~s}^{-1}$ depending on the chosen value for the magnetic field) that yields post-shock conditions matching remarkably well our observations for the Cetus Ripple components in terms of velocity, column density and velocity dispersion (see Gry \& Jenkins 2014 for details).  The origin of this shock seems compatible with a sudden increase in thermal pressure by the surrounding gas present in the Local Bubble. 
\\
\\
{\it 5.5 Possible origins for the remaining components }\\
Closer than 50 pc, the remaining components which are due neither to  the  local cloud nor to the Cetus Ripple constitute a minor fraction of the matter. They may arise from secondary shocks or small velocity waves induced by turbulent motions, or from gas escaping the local cloud. Larger components  likely to arise from separate clouds seem to exist beyond 50 pc.
\section{Summary and conclusions}
We have shown that the UV absorption line database is compatible with the picture of the LISM as one single, monolithic cloud that surrounds the Sun in all directions.

\noindent Its extent is within 9.3$\pm$7.6 pc in most directions, and 
the Sun is not close to the edge of the cloud.

\noindent Motions within the cloud deviate from that of a homogeneous rigid body. The second-order kinematics inside the cloud suggests that it may be deforming from an apparent progressive deceleration in the direction of motion. 

\noindent H I densities within the cloud vary between  0.03  and $0.1\,{\rm cm}^{-3}$, and 
the  metal abundances  show a gradient in the direction of motion.

\noindent This Local Cloud dominates the absorption in the first 50 pc. Most secondary absorption components detected in 60\% of the sight lines appear to be located inside the boundary of the Local Cloud. 

\noindent Half of the secondary components have a velocity shift of $-7.2\,{\rm km~s}^{-1}$ relative to the Local Cloud and may result from an imploding shock created by an imbalance of pressure with the surrounding medium.
\section*{References}
\begin{thereferences}
\item Frisch, P.C., Refield S. \& Slavin J.D., 2011, Annu. Rev. Astron. Astrophys. 2011. 49, 237
\item Gloeckler, G., Fisk, L. A., Geiss, J., et al. 2009, Space Sci. Rev., 143, 163
\item Gry, C. \& Jenkins, E.B., 2014, A\&A 567, A58
\item Jenkins, E. B. 2004, in Carnegie Obs Ap Serie, Vol. 4, Origin
and Evolution of the Elements, ed. A. McWilliam \& M. Rauch
\item Kimura, H., Mann, I. \& Jessberger, E. K. 2003, ApJ 582, 846
\item Lallement, R. \& Bertin, P., 1992, A\&A 266, 479
\item Malamut, C., Redfield, S., Linsky, J.L., Wood, B. E., and Ayres, T. R. 2014, ApJ 787, 75
\item McComas, D. J., Alexashov, D.,  Bzowski, M., et al, 2012, Science 336, 1291 
\item M\"obius, E., Bzowski, M., Chalov, S., 2004, A\&A 436, 897
\item Redfield, S. \&  Linsky, J.L., 2002, ApJS 139, 439
\item Redfield, S. \&  Linsky, J.L., 2008, ApJ 673, 283 
\item Wood, B.E., Redfield, S., Linsky, J.L., Mueller, H.R. \&  Zank, G.P., 2005, ApJS 159, 118
 \item Wood, B.E. \& Linsky, J. L., 1998, ApJ, 492, 788
\end{thereferences}

\medskip
\noindent A more comprehensive description of this work can be found in Gry \& Jenkins (2014).
\end{document}